\documentclass[conference]{IEEEtran}

\ifCLASSINFOpdf
\else
\fi

\usepackage{epsfig,makeidx,color,mathtools,bigints,graphicx,amsbsy,amsmath,amssymb,euscript,verbatim}
\usepackage{url}
\DeclareMathOperator{\argmin}{argmin}

\DeclareMathOperator{\tr}{\text{tr}}

\DeclareMathOperator{\vol}{vol}
\DeclareMathOperator{\vect}{vec}
\DeclareMathOperator{\diag}{diag}
\DeclareMathOperator{\rank}{rank}



\newtheorem{prop}{Proposition}

\newcommand{\Z}{\mathbf{Z}}
\newcommand{\R}{\mathbf{R}}

\newcommand{\C}{\mathbf{C}}

\begin{document}
%
\title{Nested Lattice Codes for Vector Perturbation Systems}

\author{\IEEEauthorblockN{David A.\ Karpuk, Amaro Barreal, Oliver W.\ Gnilke, and Camilla Hollanti }
\IEEEauthorblockA{Department of Mathematics and Systems Analysis\\
Aalto University\\
FI-00076 Aalto, Finland\\
Email: \{david.karpuk, amaro.barreal, oliver.gnilke, camilla.hollanti\}@aalto.fi}
}

\maketitle

\begin{abstract}
Vector perturbation is an encoding method for broadcast channels in which the transmitter solves a shortest vector problem in a lattice to create a perturbation vector, which is then added to the data before transmission.  In this work, we introduce nested lattice codes into vector perturbation systems, resulting in a strategy which we deem \emph{matrix perturbation}.  We propose design criteria for the nested lattice codes, and show empirically that lattices satisfying these design criteria can improve the performance of vector perturbation systems.  The resulting design criteria are the same as those recently proposed for the Compute-and-Forward protocol.
\end{abstract}


%
\IEEEpeerreviewmaketitle

\section{Introduction}


\subsection{Channel Pre-Inversion and Vector Perturbation}

We consider the following broadcast channel problem.  Suppose a basestation with $M$ transmit antennas wishes to transmit to $K$ non-cooperating single-antenna receivers, in the presence of fading and noise.  We assume $M\geq K$.  Assuming perfect channel state information at the transmitter, we may pre-process the data by multiplying it by the inverse of the channel matrix.  However, given some transmit power constraint, the transmitter  must rescale by a power renormalization constant, which if large can substantially affect transmission.

In \cite{swindlehurst}, it was observed that when $M = K$, multiplying the data intended for transmission by $H^{-1}$, where $H$ is the channel matrix, performs poorly in a Rayleigh fading environment as the capacity does not scale linearly with the number of users.  The authors proposed pre-multiplying instead by a regularized inverse of $H$, which causes the capacity of the resulting system to scale linearly with the number of users, but still leaves a large gap to the broadcast channel capacity.

In \cite{swindlehurst2} the authors improved on \cite{swindlehurst} using the method of \emph{vector perturbation}, in which a vector $u$ of quadrature amplitude modulation (QAM) symbols, scaled to be in the Voronoi cell of $\Z[i]^K$, is pre-processed by solving for
\begin{equation}\label{solve_for_perturbation}
x = \underset{x'\in \Z[i]^K}{\argmin}\ ||H_{\text{ZF}}(u+x')||^2,
\end{equation}
where $H_{\text{ZF}} = H^\dagger(HH^\dagger)^{-1}$ is the zero-forcing inverse of the channel matrix (a regularized inverse can similarly be used).  The transmitter then sends the vector $H_{\text{ZF}}(u+x)$, where $x$ is known as the \emph{perturbation vector}.  To remove the perturbation vector, the receivers each reduce modulo the lattice $\Z[i]$ and then decode as usual.  The performance of the system is then largely determined by the power renormalization constant
\begin{equation}
\gamma = \mathbf{E}_u||H_{\text{ZF}}(u+x)||^2
\end{equation}
which has been studied extensively, see \cite{heath_se}.  Other authors \cite{vp_lll} have studied the effect of sub-maximum-likelihood (ML) methods for computing (\ref{solve_for_perturbation}) on system performance, as well as vector perturbation methods when the users have more than one receive antenna \cite{vp_bd}.

\subsection{Summary of Main Contributions}

As far as the authors are aware, there has been no attempt to use any lattice other than the square lattice $\Z[i]$ when solving for the offset vector $x$ as in (\ref{solve_for_perturbation}).  However, the vector perturbation system model naturally generalizes to one wherein the data vectors $u$ are selected from the Voronoi cell of some complex lattice $\Lambda\subset \C^T$, and the offset vectors are selected from $\Lambda$ itself.  This naturally allows the users to employ (complex versions of) \emph{nested lattice codes}, which are known to achieve channel capacity in the additive white gaussian noise (AWGN) channel \cite{lattice_coding_awgn}.

This work represents a first attempt at introducing lattice coding into systems which employ vector perturbation.  The perturbation vector is naturally replaced by a matrix, hence we refer to our method as \emph{matrix perturbation}.  Our ultimate goal is to optimize system performance by establishing optimal nested lattice codes.  Our main contributions are as follows:
\begin{itemize}
\item In Section \ref{model}, we generalize the vector perturbation system model to one which employs nested lattice codes, and describe the matrix perturbation method.
\item In Section \ref{design_criteria}, we propose design criteria for both the fine and coarse lattices used in matrix perturbation, by studying the resulting pairwise error probability.  To this end, we employ a version of the LLL lattice reduction algorithm for complex lattices over Euclidean rings.  Interestingly, the proposed design criteria are identical to those proposed in \cite{feng_silva} for the Compute-and-Forward protocol.
\item In Section \ref{simulation_results}, we confirm the validity of our proposed design criteria when $T = 1$ by plotting the pairwise error probability of the system.
\item In Section \ref{conclusions} we conclude and discuss future work.
\end{itemize}

\subsection{Conventions}

If $A$ is a matrix with coefficients in $\C$, then $A^t$ denotes the transpose of $A$ and $A^\dagger$ the conjugate transpose of $A$.  The norm $||A||_F$ is the Frobenius norm of $A$, defined by $||A||^2_F = \tr(A^\dagger A)$.  If $A_1,\ldots,A_K$ are matrices, then $\diag(A_1,\ldots,A_K)$ denotes the block diagonal matrix with $A_k$ in the $k^{th}$ block.  If $A = (a_{ij}) \in \C^{M\times K}$ and $B\in \C^{N\times L}$, then the \emph{tensor} or \emph{Kronecker product} of $A$ and $B$ is the block matrix $A\otimes B = (a_{ij}B)\in \C^{MN\times KL}$.  If $A\in \C^{M\times K}$ then $\vect(A)\in \C^{MK\times 1}$ denotes the vectorization of $A$, given by stacking the columns of $A$ on top of each other.

\section{Matrix Perturbation System Model}\label{model}

In this section, we generalize the vector perturbation system model of \cite{swindlehurst2} to allow the users to employ physical-layer coding over $T$ time instances.  We then describe the codebooks we consider, which come from nested lattice codes.  When $T = 1$ our model specifies to the commonly-used vector perturbation model of \cite{swindlehurst2}.

\subsection{Basic Setup}\label{basic_setup}
We consider multiple-input multiple-output (MIMO) systems with $M$ transmit antennas transmitting to $K$ non-cooperating single-antenna receivers.  We model the system at time $t=1,\ldots,T$ by the equation
\begin{equation}\label{basic_model}
y(t) = H(t)s(t) + w(t)
\end{equation}
where at time $t$,
\begin{itemize}
\item $s(t)\in \C^{M\times 1}$ is the encoded data vector for transmission,
\item $H(t)\in \C^{K\times M}$ is the channel matrix, whose entries are i.i.d.\ zero-mean standard Gaussian random variables with variance $1$ per complex dimension,
\item $w(t)\in \C^{K\times 1}$ is an additive noise vector, whose entries $w_k(t)$ are i.i.d.\ zero-mean standard Gaussian random variables with variance $\sigma^2$ per complex dimension,
\item $y(t)\in \C^{K\times 1}$ is the total received vector observed, whose $k^{th}$ entry $y_k(t)$ is observed by receiver $k$.
\end{itemize}

From now on we assume a quasi-static fading model wherein $H = H(1) = \cdots = H(T)$, and  we collect the various values of $s(t)$ as columns in a matrix $S$, defined by $S = [s(1)\ \cdots\ s(T)] \in \C^{M\times T}$. Similarly we define $K\times T$ matrices $Y = [y(1)\ \cdots\ y(T)]$ and $W = [w(1)\ \cdots\ w(T)]$.  The channel equation becomes
\begin{equation}
Y = HS + W.
\end{equation}
To ensure for fair comparison over coding strategies which code over time intervals of varying lengths $T$, we normalize the transmitted signal $S$ so that
\begin{equation}\label{power_constraint}
\mathbf{E}(||S||^2_F) = \sum_{t = 1}^T\mathbf{E}(||s(t)||^2) = T.
\end{equation}
We note that $S$ can depend on $H$, and this expectation is taken over all possible $S$ for a fixed channel matrix.


We construct the encoded signal $S$ as follows.  We assume that the intended data for receiver $k$ at time $t$ is modeled by a zero-mean, uniform, discrete random variable $u_k(t)$, which are independent with respect to the index $k$.  We collect the uncoded data in a matrix $U$, defined by
\begin{align}\label{data_matrix}
U &= \begin{bmatrix} u_1 \\ \vdots \\ u_K \end{bmatrix} = \begin{bmatrix}
u_1(1) & \cdots & u_1(T)\\
\vdots & \ddots & \vdots \\
u_K(1) & \cdots & u_K(T)
\end{bmatrix} \in \C^{K\times T}, \\
 u_k &= \begin{bmatrix} u_k(1)& \cdots & u_k(T)\end{bmatrix}\in \C^{1\times T}.
\end{align}
We assume that the transmitter has perfect knowledge of the channel matrix $H$.  The transmitter constructs the encoded data matrix $S$ by computing some precoding matrix $A \in \C^{M\times K}$ which depends on $H$, and a perturbation matrix $X\in \C^{K\times T}$ (whose structure we will clarify shortly), and setting
\begin{equation}
S = A(U+X)/\sqrt{\gamma}
\end{equation}
where the power renormalization constant $\gamma$ is defined by
\begin{equation}\label{gamma_defn}
\gamma = \frac{1}{T}\mathbf{E}_U||A(U+X)||^2_F 
\end{equation}
so that (\ref{power_constraint}) is satisfied.  We assume $\gamma$ is known to all receivers.  



\subsection{Lattices}

Let $\mathcal{O}\subset \C$ be a discrete Euclidean ring, such that $\rank_{\Z}(\mathcal{O}) = 2$.  The main examples we will be interested in are the Gaussian integers $\mathcal{O} = \Z[i]$ and the Eisenstein integers $\mathcal{O} = \Z[\omega]$, where $\omega = \frac{-1+\sqrt{-3}}{2}$.

By an $\mathcal{O}$-\emph{lattice} (or simply \emph{lattice} if $\mathcal{O}$ is understood) we will mean a discrete $\mathcal{O}$-module $\Lambda \subset \C^T$.  The \emph{rank} $r$ of the lattice is its rank as an $\mathcal{O}$-module, and by the discreteness condition we have $r\leq T$.  Since $\mathcal{O}$ is a Euclidean ring, any $\mathcal{O}$-lattice $\Lambda$ of rank $r$ can be written as
\begin{equation}
\Lambda = \{x = Gz\in \C^{T\times 1}\ |\ z\in \mathcal{O}^{r\times 1}\}
\end{equation}
for a full rank matrix $G\in \C^{T\times r}$, called a \emph{generator matrix} of $\Lambda$.  The columns of $G$ form an $\mathcal{O}$-basis for $\Lambda$.  We say that $\Lambda$ is \emph{full rank} if $r = T$.  For example, the hexagonal lattice $A_2\subset\C$ can be viewed as a one-dimensional $\mathcal{O}$-lattice with $G = 1$ where $\mathcal{O}$ is the Eisenstein integers.



For any $\mathcal{O}$-lattice $\Lambda\subset \C^T$ with generator matrix $G\in \C^{T\times r}$, let $\Lambda_{\C}=\{Gz\ |\ z\in \C^{r\times 1}\}$.   Thus $\Lambda_{\C}\subseteq \C^T$ is a subspace of complex dimension $r$ containing $\Lambda$, and $\Lambda_{\C} = \C^T$ if and only if $\Lambda$ is full rank.  The \emph{Voronoi cell} of $\Lambda$ is the set
\begin{equation}
\mathcal{V}_\Lambda = \{x\in \Lambda_{\C}\ |\ ||x||^2< ||x-y||^2\ \text{for all $y\in \Lambda$, $y\neq 0$}\}.\nonumber
\end{equation}
which is a compact subset of $\Lambda_{\C}$.

We define \emph{reduction modulo $\Lambda$} for any $x \in \Lambda_{\C}$ to be 
\begin{equation}
x\ (\text{mod } \Lambda) = x - Q_{\Lambda}(x) \in \mathcal{V}_{\Lambda}
\end{equation}
where $Q_{\Lambda}(x)$ is the closest lattice point to $x$.  Thus reduction modulo $\Lambda$ sends every point $x\in \Lambda_{\C}$ to the unique representative modulo $\Lambda$ in the Voronoi cell of $\Lambda$.

For any lattice $\Lambda$, we define
\begin{equation}
r(\Lambda) = \frac{1}{2}\min_{\substack{x\in \Lambda \\ x\neq 0}} ||u||,\quad \hat{\tau}(\Lambda) = |\{x\in \Lambda\ |\ ||x|| = 2r(\Lambda)\}|\nonumber
\end{equation}
to be, respectively, the \emph{sphere packing radius} and the number of shortest vectors of $\Lambda$.  The \emph{volume} of a lattice $\Lambda$ is defined to be $\vol(\Lambda) := \vol(\mathcal{V}_\Lambda)$, and the \emph{per-dimension second moment} of a lattice $\Lambda\subset \C^T$ is defined to be
\begin{equation}
\sigma^2(\Lambda) = \frac{1}{T}\frac{1}{\vol(\Lambda)}\int_{\mathcal{V}_{\Lambda}}||z||^2dz
\end{equation}
The compactness of $\mathcal{V}_{\Lambda}$ implies that $\sigma^2(\Lambda)$ is well-defined for all $\Lambda$.  If $c\in \C$ is a constant, then $\sigma^2(c\Lambda) = |c|^2\sigma^2(\Lambda)$. If $u$ is uniformly distributed on $\mathcal{V}_{\Lambda}$, then $\sigma^2(\Lambda) = \frac{1}{T}\mathbf{E}_u||u||^2$ 

If we have lattices $\Lambda_k\subset\C^{T_k}$ for $k = 1,\ldots,K$ then we define their \emph{direct product} to be the lattice
\begin{equation}
\prod_{k = 1}^K\Lambda_k = \{[x_1^t,\ldots,x_K^t]^t\in \C^{(\sum_k T_k)\times 1}\ |\ x_k \in \Lambda_k\}
\end{equation}
for which a generator matrix is $\diag(G_1,\ldots,G_K)$, where $G_k$ generates $\Lambda_k$.  
It follows easily from the definition of the Voronoi cell that $\mathcal{V}_{\prod_{i = 1}^K\Lambda_i} = \prod_{i = 1}^K\mathcal{V}_{\Lambda_i}$.

\begin{prop}
Suppose that $\Lambda = \prod_{k = 1}^K\Lambda_k$ is the direct product of the lattices $\Lambda_k$, each of which has rank $r_k$.  Then
\begin{equation}
\sigma^2(\Lambda) = \frac{1}{\sum_{k = 1}^Kr_k}\sum_{k = 1}^Kr_k\sigma^2(\Lambda_k).
\end{equation}
\end{prop}
\begin{IEEEproof}
We omit a full proof due to length constraints, but the proposition is easily proven via direct integration when $K = 2$, after which it follows by induction for general $K$.
\end{IEEEproof}

\subsection{Encoding the Data - Matrix Perturbation}

Our approach to lattice coding roughly follows that of \cite{lattice_coding_awgn}, wherein the authors show how to use nested lattice codes to achieve the capacity of the AWGN channel.  For each user $k = 1,\ldots,K$, we assign a pair of full-rank nested lattices $\Lambda_i\subset\Lambda_i'\subset \C^T$ and define the constellation for user $i$ to be
\begin{equation}
\mathcal{C}_k = (\Lambda_k'-s_k)\cap \mathcal{V}_{\Lambda_k},\quad s_k = \mathbf{E}(\Lambda_k'\cap \mathcal{V}_{\Lambda_k})
\end{equation}
Here we have shifted by $s_k$ simply to force $\mathcal{C}_k$ to be zero-mean, allowing us to construct standard QAM constellations as such $\mathcal{C}_k$.  We will refer to $\mathcal{C}_k$ as a \emph{nested lattice code}. 

We can now make precise the nature of the perturbation matrix $X\in \C^{K\times T}$.  For a precoding matrix $A$ and a data matrix $U$ as in (\ref{data_matrix}) with $u_k \in \mathcal{V}_{\Lambda_k}\subset\C^T$, we set
\begin{equation}\label{x_defn}
X = \underset{X'\in \prod_{k = 1}^K\Lambda_k}{\argmin} ||A(U+X')||^2_F 
\end{equation}
where we view points $X'$ in the lattice $\prod_{k = 1}^K\Lambda_k$ as matrices of the form 
\begin{equation}
X' = \begin{bmatrix} x_1 \\ \vdots \\ x_K \end{bmatrix} = \begin{bmatrix} x_1(1) & \cdots & x_1(T) \\ \vdots & \ddots & \vdots \\ x_K(1) & \cdots & x_K(T) \end{bmatrix},\quad x_k \in \Lambda_k.
\end{equation}
When $T = 1$ and $\mathcal{O} = \Lambda_k = \Z[i]$ for all $k$, this is the vector perturbation strategy of \cite{swindlehurst2}, where the fine lattice $\Lambda_k'$ defines a scaled QAM constellation within the Voronoi cell of $\Lambda_k$.

Let us now fix $A = H_{\text{ZF}} = H^\dagger(HH^\dagger)^{-1}$.  The transmitter sends $A(U+X)/\sqrt{\gamma}$, in which case the observation at the receiver is
\begin{equation}
Y = HA(U+X)/\sqrt{\gamma} + W = U/\sqrt{\gamma} + X/\sqrt{\gamma} + W.
\end{equation}
Receiver $k$ observes the $k^{th}$ row of this matrix, given by
\begin{equation}
y_k = u_k/\sqrt{\gamma} + x_k/\sqrt{\gamma} + w_k
\end{equation}
at which point they multiply the above by the constant $\sqrt{\gamma}$ to arrive at the equivalent observation
\begin{equation}\label{rec_obs}
y_k' = u_k + x_k + \sqrt{\gamma}w_k.
\end{equation}
Receiver $k$ obtains the ML estimate $\hat{u}_k$ of $u_k$ from (\ref{rec_obs}) by first computing 
\begin{equation}
\tilde{y}_k = y_k'\ (\text{mod }\Lambda_k) ,\quad \tilde{y}_k\in \mathcal{V}_{\Lambda_k}
\end{equation} 
to remove the offset vector $x_k\in \Lambda_k$, and then computing
\begin{equation}\label{ML_est}
\hat{u}_k = \underset{u_k'\in \mathcal{C}_k}{\argmin}\ ||\tilde{y}_k - u_k'||^2
\end{equation}
Our goal now is to extract design criteria for the nested lattices $\Lambda_k\subset \Lambda_k'$ by studying the pairwise error probability (PEP), that is, $P(\hat{u}_k\neq u_k)$.

\section{Lattice Design Criteria}\label{design_criteria}

\subsection{PEP Analysis and Fine Lattice Design Criteria}


Let us fix a receiver $k$ and a channel $H$, and consider equation (\ref{rec_obs}).  The ML estimate $\hat{u}_k\in \mathcal{C}_k$ in (\ref{ML_est}) of the transmitted lattice point $u_k$ can alternately be described by
\begin{equation}
\hat{u}_k = \tilde{u}_k\ (\text{mod } \Lambda_k),\quad \tilde{u}_k = \underset{u_k'\in s_k+\Lambda_k'}{\argmin}\ ||y_k'-u_k'||^2
\end{equation}
where $y_i'$ is as in (\ref{rec_obs}). In essence, the reduction modulo $\Lambda_k$ receiver employed by user $k$ effectively extends the codebook $\mathcal{C}_k$ to the entire translated lattice $s_k+\Lambda_k'$.  Hence the receiver can first perform na\"ive lattice decoding in $s_k+\Lambda_k'$ to decode $\tilde{u}_k$.  The final result $\hat{u}_k$ is obtained by reducing this modulo $\Lambda_k$ to determine its equivalence class in $\mathcal{C}_k$.  

Since $\hat{u}_k\neq u_k$ implies $\tilde{u}_k\neq u_k$, we have
\begin{equation}
P(\hat{u}_k\neq u_k) \leq P(\tilde{u}_k \neq u_k) = P(\sqrt{\gamma}w_k\not\in \mathcal{V}_{\Lambda_k'}).
\end{equation}
We follow a standard union bound argument \cite[\S3.1.3]{conway_sloane}, omitting the details as the argument is so pervasive in the literature.  Letting $v_1,\ldots,v_s$ be the relevant vectors of $\Lambda_k'$ and setting $r_j = ||v_j||/2$, the union and Chernoff bounds yield
\begin{equation}\label{PEP_upper_bound}
P(\sqrt{\gamma}w_k\not\in \mathcal{V}_{\Lambda_k'}) \leq \sum_{j = 1}^s e^{-r_j^2/\gamma\sigma^2}.
\end{equation}
Considering the largest summands in (\ref{PEP_upper_bound}) yields the approximate upper bound
\begin{equation}\label{loose_PEP_upper_bound}
P(\hat{u}_k\neq u_k) \lesssim \hat{\tau}(\Lambda_k')e^{-r(\Lambda_k')^2/\gamma\sigma^2}
\end{equation}
where $r(\Lambda_k')$ is the sphere packing radius of $\Lambda_k'$ and $\hat{\tau}(\Lambda_k')$ the number of minimal vectors in $\Lambda_k$.  Assuming that $\gamma$ is relatively insensitive to the choice of fine lattice, we see that the optimal $\Lambda'_k$ are those which are good for the AWGN channel.  Furthermore, from (\ref{loose_PEP_upper_bound}) we see that the nested lattice code should be chosen to minimize $\gamma$.  

\subsection{Analysis of $\gamma$}

From the estimate (\ref{loose_PEP_upper_bound}) we see that a full analysis of the PEP requires us to study how the power renormalization constant $\gamma$ varies with the nested lattice code.  Following an argument of \cite{heath_se}, we show in this section that it can be approximated (up to a factor of $K$) by the second moment of a lattice.

Recalling the definition of $\gamma$ from (\ref{gamma_defn}) and using basic facts about Kronecker products and vectorization yields
\begin{equation}
\gamma = \frac{1}{T}\mathbf{E}_U||(A\otimes I_T)\vect((U+X)^t)||^2,
\end{equation}
where for a given $U$, the perturbation matrix $X$ is chosen among all $X'\in \prod_{k=1}^K \Lambda_k$ to minimize this quantity.  

Let us now consider the $\mathcal{O}$-lattice
\begin{equation}\label{big_L}
\mathcal{L} = (A\otimes I_T)\prod_{k = 1}^K\Lambda_k \subset \C^{MT}
\end{equation}
which has rank $KT$ and generator matrix
\begin{align}\label{complicated_tensor}
G_{\mathcal{L}} &= (A\otimes I_T)\diag(G_1,\ldots,G_K) \\
&= \begin{bmatrix} A^{(1)}\otimes G_1\ \cdots\ A^{(K)}\otimes G_K\end{bmatrix}
\end{align}
where $A^{(k)}$ is the $k^{th}$ column of $A$.  In particular when all users employ the same coarse lattice $\Lambda$ with generator matrix $G$, the generator matrix of $\mathcal{L}$ is given by $G_{\mathcal{L}} = A\otimes G$.

As the columns of $U^t$ corresponds to elements of the various codebooks $\mathcal{C}_k = \Lambda_k'\cap \mathcal{V}_{\Lambda_k}$, we have
\begin{equation}
(A\otimes I_T)\vect(U^t) \in (A\otimes I_T)\prod_{k = 1}^K\mathcal{V}_{\Lambda_k} = (A\otimes I_T)\mathcal{V}_{\prod_{k = 1}^K\Lambda_k}\nonumber
\end{equation}
and $(A\otimes I_T)\vect(X'^t) \in \mathcal{L}$ for any $X'\in \prod_{k = 1}^K\Lambda_k$.    Following the argument of \cite[Lemma 1]{heath_se}, it follows from the definition of $X$ (the optimal such $X'$) that
\begin{equation}
(A\otimes I_T)\vect((U+X)^t) \in \mathcal{V}_{\mathcal{L}}.
\end{equation}

Now let us approximate the distribution of $\vect(U^t)$ by the uniform distribution on $\prod_{k = 1}^K\mathcal{V}_{\Lambda_k} = \mathcal{V}_{\prod_{k = 1}^K\Lambda_i}$.  It follows from the above that $(A\otimes I_T)\vect((U+X)^t)$ is approximately uniformly distributed on $\mathcal{V}_{\mathcal{L}}$, in which case $\gamma$ is approximated as follows:
\begin{align}
\gamma &= \frac{1}{T}\mathbf{E}_U||(A\otimes I_T)\vect((U+X)^t)||^2 \\
&\approx  \frac{1}{T}\frac{1}{\vol(\mathcal{V}_\mathcal{L})}\int_{\mathcal{V}_{\mathcal{L}}}||z||^2\ dz =K\sigma^2(\mathcal{L}) 
\end{align}
from which it follows that for a fixed channel $H$, the coarse lattices $\Lambda_1,\ldots,\Lambda_K$ should be chosen to minimize the second moment $\sigma^2(\mathcal{L})$.  In the next subsection we propose an approximation of $\sigma^2(\mathcal{L})$ which clarifies how $\sigma^2(\mathcal{L})$ depends on the various coarse lattices $\Lambda_k$.


\subsection{Coarse Lattice Design Criteria}

Recall that the LLL algorithm \cite{LLL_original} takes as input an integer basis of a $\Z$-lattice and outputs an \emph{LLL-reduced} basis, with the property that the basis vectors are in some sense as orthogonal as possible.  A variant of the LLL algorithm introduced in \cite{LLL_euclidean} generalizes the idea of an LLL-reduced basis to $\mathcal{O}$-lattices, where $\mathcal{O}$ is any Euclidean ring.

Let $\Lambda\subset\C^M$ be an $\mathcal{O}$-lattice of rank $K$ and let $A$ be its generator matrix, whose columns form an $\mathcal{O}$-basis for $\Lambda$.  The output of the LLL algorithm of \cite{LLL_euclidean} when run on $\Lambda$ can be viewed as a matrix decomposition of the form
\begin{equation}
A = BZ
\end{equation}
where the columns of $B$ form an $\mathcal{O}$-LLL reduced basis (see \cite{LLL_euclidean}) for $\Lambda$ and $Z\in \mathcal{O}^{K\times K}$ is unimodular, meaning that $|\det(Z)| = 1$ and $Z^{-1}\in \mathcal{O}^{K\times K}$.  From the unimodularity of $Z$ it follows that $B$ generates the same $\mathcal{O}$-lattice as $A$.  

Let $B = QR$ be a QR-decomposition of the $\mathcal{O}$-LLL reduced generator matrix $B$ of $\Lambda$.  Since $R$ is both upper-right triangular and `almost' orthogonal, the off-diagonal entries of $R$ are close to zero.  We thus approximate $R$ by the diagonal matrix
\begin{equation}
R\approx R_0,\quad R_{0,ij} = \left\{\begin{array}{cl}
r_{ii} & i = j \\
0 & i\neq j
\end{array}\right.
\end{equation}
which simply sets all off-diagonal entries of $R$ to zero.  Let us now set $B_0 = QR_0$.

Consider now the $\mathcal{O}$-lattice $\mathcal{L} = (A\otimes I_T)\prod_{k = 1}^K\Lambda_k$ as in (\ref{big_L}), whose per-dimension second moment approximates the power renormalization constant $\gamma$.  Let $\mathcal{L}_0$ be the $\mathcal{O}$-lattice  $(B_0\otimes I_T)\prod_{k = 1}^K\Lambda_k$, where $B_0$ is obtained from $A$ by the above-outlined procedure.  We approximate $\gamma$ as follows:
\begin{align}
\gamma &\approx \sigma^2(\mathcal{L}) \approx \sigma^2(\mathcal{L}_0) \\
&= \sigma^2((B_0\otimes I_T)\prod_{k = 1}^K\Lambda_k) = \sigma^2((R_0\otimes I_T)\prod_{k = 1}^K\Lambda_k) \\
&= \frac{1}{K}\sum_{k = 1}^K|r_{kk}|^2\sigma^2(\Lambda_k)
\end{align}
From the above we conclude that the coarse lattices should be chosen to minimize $\sigma^2(\Lambda_k)$, that is, they should be good for quantization.

\subsection{A Connection to Compute-and-Forward}
Summarizing the design criteria derived in the previous three subsections, we see that the nested lattice codes $\Lambda_k\subset\Lambda_k'\subset\C^T$should be chosen so that:
\begin{itemize}
\item[(i)] $\Lambda_k'$ is good for the AWGN channel, and
\item[(ii)] $\Lambda_k$ is good for quantization.
\end{itemize}
Lattice coding has also been proposed for the Compute-and-Forward (CaF) protocol \cite{nazer_gastpar} for relay networks.  An algebraic approach to CaF was taken in \cite{feng_silva} in which the authors use the PEP to extract design criteria.  Interestingly, the nested lattice code design criteria proposed in \cite{feng_silva} are identical to the design criteria derived above for the matrix perturbation technique.  While we will not pursue this connection in this paper, it certainly merits further investigation.

\section{Simulation Results}\label{simulation_results}

We now present first simulation results which confirm the legitimacy of our design criteria for lattices $\Lambda \subset \C$, so that $T = 1$.  We compared the Gaussian lattice $\Z[i]$ (i.e.\ QAM modulation) which is commonly used in vector perturbation with the hexagonal lattice $A_2$, which is both a better lattice for the AWGN channel and a better quantizer than the Gaussian lattice.  For each value of $K$ we sampled $10^3$ channel matrices $H$, and for each $H$ we simulated the transmission of $10^3$ data vectors $u$ at each value of $1/\sigma^2$.  For a fixed $\Lambda$ the value of $\gamma$ apparently does not vary much with $H$, and hence accurate error results can be obtained with a somewhat small number of channels as the only effect of $H$ is on $\gamma$.

\begin{figure}
\centering
\includegraphics[width=.35\textwidth]{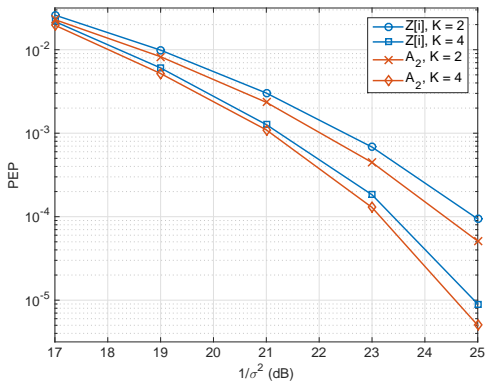}
\caption{PEP for user $k = 1$, in vector perturbation systems with $K = M = 2$ and $K = M = 4$, when both users employ the same fine lattice $\Lambda'$ and the same coarse lattice $\Lambda = 2^4\Lambda'$.  Here we compared the Gaussian lattice $\Z[i]$ commonly used in vector perturbation with the hexagonal lattice $A_2$.}
\label{Zi_vs_A2_K24M24_N16}
\end{figure}

\begin{figure}
\centering
\includegraphics[width=.35\textwidth]{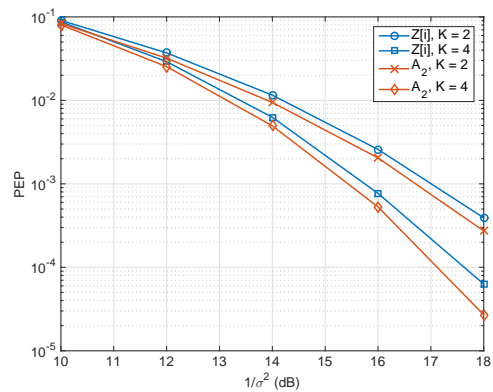}
\caption{The same simulation parameters were used as in Fig.\ 1, but with $M = 3K/2$ for all systems.}
\label{Zi_vs_A2_K24M36_N16}
\end{figure}

In Fig.\ \ref{Zi_vs_A2_K24M24_N16} we plot the PEP for user $k = 1$, in vector perturbation systems with $K = M = 2$ and $K = M = 4$, when both users employ the same fine lattice $\Lambda'$ and the same coarse lattice $\Lambda = 2^4\Lambda'$.  When $\Lambda' = \Z[i]$, this is equivalent to standard vector perturbation \cite{swindlehurst2} with $16$-QAM modulation.  In Fig.\ \ref{Zi_vs_A2_K24M36_N16} we repeat the experiment for systems with $K= 2$ and $K = 4$ with $M = 3K/2$.

In Fig.\ \ref{Zi_vs_A2_K24M24_N16} we see that using the lattice $A_2$ improves the performance of standard vector perturbation techniques by about $0.5$ dB at higher values of $1/\sigma^2$, for both system sizes.  Note that system performance apparently increases with $K$; this is due to the fact that $\gamma$ decreases with $K$, though it quickly approaches a constant value (see Fig.\ 1 of \cite{heath_se}).  We see from Fig.\ \ref{Zi_vs_A2_K24M36_N16} that similar results are obtained when $M = 3K/2$.  These preliminary simulation results only treat the case of $T = 1$, though in analogy with traditional lattice coding \cite{lattice_coding_awgn} we expect using higher dimensional lattices (i.e.\ $T>1$) will yield further improvements in system performance.

\section{Conclusions}\label{conclusions}

In this paper we investigated the use of nested lattice codes in systems employing vector perturbation for broadcast channels.  Design criteria based on the PEP were proposed for nested lattice codebooks, and it was observed that the fine lattice should be good for the AWGN channel, and the coarse lattice should be good for quantization.  Interestingly, these are the same proposed design criteria for CaF derived in \cite{feng_silva}.  Future work includes studying how nested lattice codes perform in conjunction with regularized inversion \cite{swindlehurst}, and generalizing to broadcast channels in which the receivers have more than one antenna, in particular to systems employing the block diagonalization technique of \cite{swindlehurst3}.





\bibliographystyle{ieee}
\bibliography{myrefs_new}

\end{document}